On an RNA-membrane protogenome

Michael Yarus

Department of Molecular, Cellular and Developmental Biology
University of Colorado  Boulder
Boulder, Colorado
Email: yarus @colorado.edu
Phone: +1 303 8176018
Web: www.colorado.edu/MCDB/people/emeritus/mike-yarus

**Abstract**

Selected ribonucleotide sequences bind well to zwitterionic phospholipid bilayer membranes, though randomized RNAs do not. There are no evident repeated sequences in selected membrane binding RNAs. This implies small and varied motifs responsible for membrane affinity. Such subsequences have been partially defined. Bound RNAs require divalents like $Mg^{2+}$ and/or $Ca^{2+}$, preferring more ordered phospholipids: gel, ripple or rafted membranes, in that order. RNAs also bind and stabilize bilayers that are bent or sharply deformed. In contrast, RNA binding without divalents extends to negatively charged membranes formed from simpler anionic phospholipids, and to plausibly prebiotic fatty acid bilayers. RNA-membranes also retain RNA function, such as base pairing, passive transport of tryptophan, specific affinity for peptide side chains, like arginine, and catalysis by ribozymic ligase. Multiple membrane-bound RNAs with biochemical functions, linked by specific base-pairing are readily constructed. Given these experimental facts, genetic effects seem plausible. RNA functions often reside in few nucleotides, and are easily joined in a small RNA. Base-paired groups of these can evolve to be purposeful, joining related RNA functions. Such RNA groups permit complex genome functions, but require only replication of short RNAs. RNA-membranes facilitate accurate RNA segregation at cell division, and quickly evolve by appending new base-paired functions. Thus, ancient RNA-membranes could act as a protogenome, supporting orderly encoded RNA expression, inheritance and evolution before DNA and the DNA genome.

**Introduction**

**Multiple lipid structures in bilayers.** Biological compartments are delimited by bilayer membranes. They have a hydrophilic exterior contacting the aqueous environment and a hydrophobic membrane interior that obstructs water and other polar molecules (Fig. 1A). Modern cell outer membranes and the limits of cell organelles contain phospholipids, which are not necessarily uniform and planar, but intermix multiple lipid conformations (1, 2).

Fluid – the least ordered of bilayer membrane forms, allowing lipid molecules to diffuse freely and change in conformation, with acyl chains isomerizing from extended to kinked.

Liquid-ordered – an intermediate membrane with freely-diffusing lipids having extended orderly acyl groups within a bilayer; these are irregularly intermixed with planar, more fluid regions.

    Ripple – some fluid lipid bilayers, when cooled, congeal into corrugated intermediate regions with some ordered and some less ordered lipid forms alternating in regular ridges (3).

    Rafted – areas of lipid that intermix sphingolipid, cholesterol and common phospholipids in a more ordered complex, occurring as islands within a more fluid bilayer (4). Rafts concentrate cellular molecules to perform essential biochemical functions.

    Gelled – near-solid condensed immobile lipids below their major melting transition, constituting a maximally ordered bilayer.

Membrane forms above are listed, top to bottom, with increasing order (5). This is useful, because RNA prefers binding to phospholipid layers in more ordered states (6). Typical experiments controllably decrease membrane order by increasing temperature, using unsaturated rather than saturated fatty acids as acyl groups, employing shorter fatty acids that support weaker hydrophobic bonds in membrane interiors, reducing $Ca^{2+}$ or increasing $Na^+$ (7).

**Selected RNAs with membrane affinity.** Selection of RNAs with membrane affinity from random-sequence molecules seemed likely to clarify requirements for RNA-membrane binding; this has proven partially true.

Free RNAs are smaller than liposomes, so easily resolved by size-exclusion chromatography (8) from first-eluted liposomes. Using liposomes with fluid phosphatidylcholine-cholesterol membranes, RNAs having 50 internal randomized nucleotides were selected for migration with liposomes. Initial randomized RNAs did not detectably bind. But after 11 cycles of selection, 70% of individual sequences stably re-bound to the exterior of the liposomes, requiring 20mM $Mg^{2+}$ and 10 mM $Ca^{2+}$. Four of 8 tested RNAs disrupted bilayers, releasing $^{22}Na^+$ from liposomes. Thus, four of 8 selected RNAs bound without disturbing membrane continuity. Selected RNAs acted on biological cell membranes, making patch-clamped HEK293 cells conductive. Truncated 53 and 44 ribonucleotide RNAs were membrane-active.

Selected RNAs foreshadowed later isolates in several ways (8): no repeated sequences were evident among selected sequences, though tracts of repeated purines and pyrimidines were frequent. Thus though membrane-binding sequences are not evident among randomized RNAs, they are apparently readily selected. Binding sequences must be (small) and multiple - even in a population whose majority binds, no repeats are recognizable. Secondly, 5 mM choline (but not homologous ethanolamine) prevents binding. RNA affinity for the outward polar headgroup of phosphatidylcholine likely has a major role in RNA binding. Thirdly, membrane bound RNAs usually stay in the leaflet first encountered (Fig. 1B); they remain outside exposed vesicles (8), exposed to exterior ions and macromolecules like protein enzymes.

The conclusion that membrane interactions are mediated by small, varied sequences is supported by multiple efforts in tracing bilayer interactions to repeated RNA motifs. These range from the above repetitive purine and pyrimidine tracts (8), to repeated AG (9), G quartets (9), to UUGU, UCCC, CUCC, CCCU (10, 11), to CCCU and GGAG (12), to UUUCU, UUCAC, UUGCAC, UUUUCC, and UCUCU (13). The sequence CAAUUCCAG is protected from digestion after membrane association (8). But conclusively membrane- bound RNAs exist that contain none of these (8). Thus, the distribution of affinity between polar headgroups and hydrophobic membrane cores, and the way that RNA sequences contact these (Fig. 1B), await final experimental definition. Such experiments will require bilayers of defined and varied structure: gel, ripple, rafts, ordered and fluid (6).

However, membrane residence can also be simple, conferred by a small hydrophobic base modification. This is exemplified by a 5-carbon isopentenyl anticodon loop modification of human tRNA$^{Sec}$, which preferentially binds rafted bilayers (14).

**Fluid membrane affinity.** Stable membrane binding selected using reduced divalents (5mM $Mg^{2+}$ and 2mM $Ca^{2+}$) does not isolate RNA monomers (15). Individual RNAs must interact to bind efficiently and alter fluid phospholipid membrane permeability (Fig. 1C). One efficient pair is RNA9 and RNA10 (Fig. 2A and 2B), which bind optimally when supplied in a 2:1 molar ratio, though weak fluid membrane affinity is detectable for RNA9 alone under these conditions. Native gel electrophoresis suggests that an RNA:9:10 kissing loop (16) complex oligomerizes further when it binds fluid phosphatidylcholine membranes stably. RNA9:RNA10 oligomers also disrupt black membranes and release interior liposome GT$^{32}$P; they therefore probably stabilize transient bilayer pores. Selection of active RNA complexes re-emphasizes simple membrane affinity sequences; selected partners must be frequent enough to reproducibly oligomerize during selection among randomized RNAs.

**RNAs bind constrained lipids.** RNA9:RNA10 membrane oligomers are confirmed by tapping-mode atomic force microscopy microscopy using flattened phosphatidylcholine vesicles supported on hydrophobic mica (5). RNA10 alone on mica appears in distributed individual monomers. RNA9 on mica is in varied chains, but if its RNA9:9 kissing loop interaction with itself (Fig. 2B) is disrupted by mutation of two nucleotides, it then appears as monomer RNAs. In contrast, mixed RNA9:RNA10 on mica is very heterogeneous, with 29 weight percent as longer complexes, including chains. Most interestingly, in the lipid vesicle layer mixed RNA9 and RNA10 oligomerize further, forming thick bands preferring the edges of lipid patches. Apparently, such RNAs prefer the unusual lipid conformation of sharply bent membranes at flattened liposome edges.

This is further emphasized by fluorescence microscopy of labeled RNA9:RNA10 RNA bound to distinctly fluorescent membranes (Fig. 2 of ref. 5). RNA appears throughout membranes, tracking membrane fluorescence accurately, but concentrating at bends. For example, particularly concentrated at the constriction between joining or separating vesicles. This predicts structural effects: affinity for curved bilayers enables RNAs to favor not only transient pores, but also membrane fusion or division (Fig. 1C).

**The RNA9:RNA10 interaction.** Selected RNA sequences (Fig. 2) rationalize membrane functions. RNA10 (Fig. 2A) is an extended 113-mer. Widely separated small sequences apparently account for its oligomerization on membranes; for example, RNA10 pairs to RNA9 via the sequence CUGCCC at nt47 (Fig. 2A), chemically protected when RNA10 pairs with RNA9. A deoxyoligonucleotide of similar sequence prevents RNA10:RNA9 interaction (15). The anteparallel RNA9 complement GGGCAG (Fig. 2B) is also protected by RNA10:RNA9 pairing. The RNA9 118-mer forms a stable dimer with itself (15) at the self-complementary tetramer GAUC at nt106 (also present near the constant 3' terminal sequence of RNA10). An overlapping deoxoligo prevents RNA9 dimerization, as does mutation of RNA9 GAUC nucleotides (15).

These observations are combined in Fig. 3, which radically simplifies RNA9 (Fig. 2B) and RNA10 (Fig. 2A) sequences as closed outlines in order to clearly show their association, which is implicated in membrane interaction (15) and strongly constrained by RNA polarities and sequences. Fig. 3's outlines, like the RNAs themselves, have polarities indicated by green arrowheads in boundary lines.

In Fig. 3A, individual RNA outlines are presented, each named with a yellow number within its larger loop; RNA9 is blue, and RNA10 red. As just above, complementary interacting sequences are presented beside larger (leftward) and smaller (rightward) kissing loops. RNA10 is shown in two forms, the leftward red form being the simpler selected RNA10 (Fig. 3A). The red outline to the right is an experimentally composed RNA10 with an inserted triangular functional domain replacing some RNA10 sequences (Fig. 3A), as observed in a passive RNA-membrane transporter that binds tryptophan (17), or adds specific arginine affinity to a bilayer (18).

In Fig. 4B and Fig. 4C, selected membrane-active RNA9:RNA10 complexes are shown, paired via loop sequences highlighted in Fig. 2A and Fig. 2B. Native gel electrophoresis (15) and atomic force microscopy (5) consistently suggest RNA9 alone forms a stable dimer with itself, while RNA10 alone exists as a monomer, probably because its GAUC sequence is engaged in secondary structure (Fig. 2A). When RNAs are mixed, different larger aggregates are formed as chains and more complex shapes (5). Size-exclusion chromatography selects stable binding to membranes because early-eluted liposome-bound RNAs must survive in the absence of unbound RNAs. Bound RNAs form optimally when a stoichiometry of 2 RNA9:1 RNA10 are supplied, and membrane-bound RNAs are recovered with the same 2 RNA9: 1 RNA10 stoichiometry (15).

Fig. 4B shows complexes that can form chains of any length by repeating an RNA10:RNA9 unit, as suggested by triple dots at Figure right. Fig. 4B explains observed long chains (5), which would have 1:1 stoichiometry.

Fig. 4C shows a complex of stoichiometry 2 RNA9:1 RNA10, which can explain recovery of this property from liposome-bound RNAs. Figure 4C uses the observed stable RNA9 dimer at both ends, pairing them via hexanucleotide kissing loops to a central RNA10 dimer. Both dimers are paired via GAUC kissing loops. Fig. 4C's structure breaks down to a stable RNA9:RNA10 dimer in the presence of the GAUC loop DNA oligo competitor, as observed (15). The Fig. 4C structure probably dominates stoichiometry of bound RNAs because, in fluid phospholipid membranes, RNA9 supplies membrane affinity, while unaltered RNA10 alone is not stable on liposomes (15). Fig. 4C is the molecular complex with the greatest RNA9:RNA10 ratio, which might well bind most quickly and be the major RNA retained after incubation on resolved fluid liposomes.

Thus RNA9:RNA10 on fluid membranes (5) retains base pairing comparable to that in solution (15).

**Bound RNAs preserve function.** RNA-membranes also execute complex functions requiring preservation of solution RNA structure. An initial example was a passive membrane transporter for tryptophan composed of RNA (17). An internal section of RNA10, which had given no biochemical indication of a membrane role (15), was replaced (Fig. 2A) with a Trp binding site (19). This modified RNA10Trp (Fig. 2A) still bound fluid liposomes when aided (Fig. 2B) by RNA9 (15). Tryptophan-binding was intact also, though with a higher $K_D$. The amino acid site was still Trp-specific, both complexed in solution, and when membrane bound. Liposome-bound RNA10Trp:RNA9 admitted the amino acid to liposomes at an enhanced rate, confirmed both isotopically, as well as by Trp FRET to an internal liposome chromophore. Increased permeability was specific, not general, and liposome fusion was ruled out by controls. Calculated transport was comparable to passive protein transporters using similar mechanisms.

Similarly, an RNA10Arg, built by similar replacement of the RNA10 interior with an RNA site for Arg (Fig. 2A), retained arginine specificity, being unresponsive to 8 other amino acids (18). Interestingly, a second internal tract in RNA10Arg (Fig. 2A) was resynthesized (doped) with 5% of the other 3 nucleotides at all positions and reselected, yielding increased membrane and Arg affinities. This included a UUG to CCCG alteration (Fig. 2B) that enabled reselected RNA10Arg to bind rafted and rippled phospholipid bilayers without RNA9 assistance.

The R3C ligase ribozyme (20) is active on membranes (9). If ribozyme ligation is inhibited by addition of a terminal membrane RNA sequence, ligase recovers its activity when the inhibitory sequence is membrane-bound.

**Bound RNAs shape membranes.** Bound RNAs take up space and are usually exclusive to the membrane leaflet that first contacts them (Fig. 1A). Bound RNAs enlarge their contacted leaflet; they tend to bend their membrane away from themselves at bound sites (Fig. 1B).

Moreover, membrane lipids have intrinsic shapes that can collaborate with RNA shaping. Membrane lipids can have small headgroups (conical lipids), or small hydrophobic acyl groups (reverse conical lipids) and so impel a bilayer leaflet to bend (21). Alternatively, headgroups and acyl groups can balance (cylindrical lipids), favoring a flat membrane. Fig. 1C depicts a bend, favored by two base-paired, bound RNAs and aided by specific lipids in membrane region A and B. In the example, large lipid headgroups in **A** and large acyl groups in **B** (Fig. 1C) would collaborate with membrane-bound RNAs to bend their membranes.

**Rafts choose RNA sequences.** RNA-membrane interaction is strongly altered by membrane lipid structure. Using phosphatidylcholine liposome membranes with different acyl groups, at different temperatures, ± cholesterol to favor lipid order, chemically similar membranes in differing structural states were compared for RNA affinity. As a rule, bound RNA increases with phospholipid layer order (6).

More specifically: the selected 113-mer membrane RNA10 (15), which stably adhered to fluid phosphatidylcholine membranes only when associated with RNA9 (15), now binds unassisted to more ordered bilayers. Both rippled (DMPC @ 18-23°) and rafted (6 DOPC: 3 sphingomyelin: 1 CHOL @ 23°) membranes bind RNA10, but release it when raised temperature makes these bilayers fluid. Because all temperatures used are well below RNA $T_m$, these are probably wholly membrane effects. Interestingly, all tested RNAs bind ≈ equivalently to rippled bilayers, but rafted bilayers make strong structural distinctions. For example, rafts bind forms of RNA10 well, but sequence-randomized RNAs very poorly. Thus, rafted bilayers are a target for structure-specific RNA affinity.

**RNA binding extends to singly-charged lipid headgroups, including fatty acids.** The above phenomena occur in complex neutral zwitterionic phospholipid membranes, prevalent in modern cells. However, interaction with RNA extends to simpler lipid bilayers, accordingly more likely during primitive cellular life.

Cationic dioleyl trimethyl ammonium propane bilayers on quartz surfaces stably bind RNA9 with RNA10 (15) under simple conditions, with only buffer and $Mg^{2+}$ present (22). This interaction is presumably principally mediated by electrostatic complementarity between negative charges on RNA and a positive membrane lipid headgroup.

Anionic, negatively charged phosphatidyl glycerol membranes bind RNA10 in buffer and NaCl (23). The membrane destabilizes the RNA to temperature, while the RNA10 simultaneously stabilizes the membrane and increases its melt cooperativity. Thus, uniform negative membrane surface charge is not a bar to RNA interactions, and (unlike zwitterionic phosphatidylcholine bilayers) divalent ions are not required for these interactions, which alter lipid properties.

Another kind of anionic extension exists for saturated 16-carbon palmitic acid, a possibly primitive (24) bilayer-forming fatty acid. Using RNA binding to liposome-coated magnetic beads (in buffer, $Mg^{2+}$ and $Ca^{2+}$ at 24°) such membranes bind randomized RNA sequences significantly better than control fluid membranes, and in fact, only ≈ four-fold less well than rippled phosphatidylcholines (9). Affinity for randomized sequences clearly suggests common, varied RNAs with affinity for palmitate membranes.

Accordingly, fatty acid-RNA interactions exist in candidate primordial membranes. Such bilayers are more fragile, but more permeable, which may serve the metabolic needs of early cells (25); for example, by improving access to ribose (26). Fatty acid bilayers are also sensitive to $Mg^{2+}$, often required for RNA activities. But such sensitivity is relieved by mild chelation of the divalent (27), or in membranes composed of varied lipid types (28). Fatty acid bilayers appear as plausible primordial boundaries below.

## Discussion

**RNAs prefer structured targets.** Much of what is known about RNA interaction with membranes appears an effect of fitting RNA to a partially structured binding site. That is, many observations suggest that the cost of rearranging lipid molecules making contact with bound RNA is highly significant. For this reason, RNAs bind bilayers best (and independent of RNA structure) if they are gelled, strongly but somewhat less if bilayers are rippled and are less well bound but discriminate RNA structures if rafted. Weakest binding occurs for fluid bilayers, where lipid rearrangements are most extensive and costly. Thus the entropic cost of fixing otherwise somewhat free lipids and the enthalpic penalties for forcing lipids into a mutually-less-favored bound environment dominate RNA affinity.

Lipid rearrangements account for the accumulation of RNAs bound to fluid membranes at sharply bent locations, like the junctions of fusing or dissociating vesicles, similarly explaining RNA9:RNA10 oligomer preference for the edges of flattened vesicles where membranes turn under sharply. In these cases, geometric deformation of the bilayer constrains its lipids, mimicking loss of lipid freedom in more ordered bilayers. One might theorize that some lipid constraints are inhibitory; but in known cases, none yet appear so. RNA prefers bilayers less free to change. However, rafts are a special case. Limited lipid adjustment with differing RNAs can be exploited: rafted phospholipid membranes choose bound RNA structures demanding the most favorable rearrangements.

Experimentally, both lipid and RNA are changed somewhat at RNA binding sites. Catalysis by R3C ligase ribozyme can be speeded or slowed on membrane association (9). RNA10 (15) alters the temperature profile of bilayer melting (6). In fluorescence microscopy, RNA10 (15) also accurately and completely coats large rafted areas on vesicles (6). Such coating is not definitive, but suggests that the altered RNA-bound lipid may stimulate fusion of smaller RNA- enclosing rafts to generate larger functional raft units, an important consideration for their activities.

There is another way to view lipid preference. RNA affinity for constrained membranes implies that RNA binding favors unusual lipid arrangements. Accordingly, membrane RNAs may facilitate many biological events (Fig. 1C): invagination / endocytosis, vesiculation / exocytosis, tubulation. Symmetrical vesiculation is otherwise termed cell division, so RNAs could induce or guide ribocyte division (29).

**Primordial size matters; modular RNAs.** RNAs only a few ribonucleotides in length exhibit complex functions. A ribozyme consisting of a tetramer and a pentamer synthesizes tetramer aminoacyl-RNA which is also converted to peptidyl-RNA (30). This short ribozyme also is active when inserted into a longer RNA (31). Competent RNAs can join several active sequences. Such RNAs partially solve the RNA replication problem (32), because fewer internucleotide bonds make replication more accessible.

Accordingly, one readily envisions (Fig. 4) short RNAs that specifically bind bilayers, also contain varied functional domains, and further, pair to build extensive paired structures (17, 18). These RNAs must also maintain functional sequences that enable their reproduction, as does RNA10. Fig. 4 relies on RNAs that express complex functions by being modular, composed of small, multifunctional, relatively easily replicated subunits.

**Rafts organize RNAs.** Blue areas (Fig. 4A) are rafted membrane regions, distinct from the major gray exterior membrane surrounding them, viewed from the cell interior. Smaller red structures are varied functional RNAs (functions **a**, **b**, **m**, **n**, **o**) also having the small sequences discussed above, required for rafted membrane affinity. Raft and functional sequences in one sequence pose no difficulty (Fig. 2, Fig. 3) because membrane affinity requires few mutations in an non-membrane RNA (15). Rafts will specifically bind these, ignoring RNAs lacking required sequences. Raft affinity suggests segregating ribocyte RNAs, alongside functional RNAs having no membrane affinity, not inherited precisely.

**Initiation of division.** Some Fig. 4 RNAs (**x**, **y**) alter lipid conformation. These are the type repeatedly observed (Fig. 1C), that promote membrane curvature. A sufficient number of such RNAs can initiate a furrow which facilitates division by expanding to transect the cell (Fig. 4C). Such events can also be aided by conical or reverse conical lipids that help form and extend a furrow (Fig. 1C). In this RNA-initiated case, division is not arbitrary but determined by RNA, like pairing between RNAs as in Fig. 1C. Such a mechanism could manifest an RNA cell cycle, dependent on cyclic availability of RNAs **x** and **y**.

Alternatively: a division furrow initiated by other cell events can collect RNAs (5). Base pairing makes segregation of RNAs during division a matter of efficient biophysical chemistry, rather than random assortment. RNA-membranes therefore emulate a later base-paired DNA genome, though requiring only documented RNA and lipid activities. RNAs joining a forming furrow suggest dual RNA roles; as free structures and catalysts and then (33, 34), after division has been initiated, as units of inheritance.

**Base-paired, rafted RNA assemblies.** RNAs add to rafted areas, guided by complementary base-pairing to an existing raft RNA. This means that Fig. 4's rafted RNAs are related groups, linked by one or more small base-pairing sequences. These options are not theoretical; such paired and unpaired elements exist in the specific paired kissing loops (Fig. 2) of RNA9 and RNA10 (5, 15). Raft groups can evolve to be functional; for example, all RNAs agents in a biochemical pathway (e.g., RNAs with **m**, **n**, and **o** functions in Fig. 4A).

Rafted RNAs are ordered for division. RNAs **y** tie groups to the furrow (Fig. 4A). Stressed furrow lipids will not be rafted; instead the furrow is hedged in on both sides by rafts. Base paired groups span the furrow due to outward directed pairing nucleotides of RNA **y**. RNAs paired to RNA **y** bind rafts, as does RNA10. Thus groups of **y**-bound RNAs will be divided as the furrow grows inward. Multiple **y** groups (Fig. 4) help insure that RNA **y** functions also are inherited.

Raft contents are not replicated in the usual sense, but instead repeatedly assembled in ordered ways guided by base pairing. Such mechanisms are accurate, but not completely reliable, and Fig. 4 depicts two potential errors. Firstly, the RNA with functional domain f in Fig. 4A binds rafts as does doped RNA10Arg (18), using independent raft affinity. However, it cannot segregate accurately.

Imprecision in membrane assemblies would also include absent RNAs. Note that the rafted RNA with **m** function, the terminal RNA of its group, can readily be lost or absent, as it is on both sides of the large rafted groups in Fig. 4A. However, this irreproducibility can be compensated statistically by employing multiple genetic groups, paralleling RNA **y** above.

**More fatty acid data.** It is not yet clear whether all Fig. 4 functions exist in simpler bilayers. Are there varied lipid structures within fatty acid membranes? Headgroups and acyl chains of varied properties suggest inhomogeneous fatty-acid bilayers could exist, but such possibilities must be experimentally explored. In any case, Fig. 4B shows RNAs on non-rafted surfaces, requiring only general membrane affinity, already observed (9). Functional RNAs **a** and **b** (Fig. 4B) pair with an RNA with membrane-distorting function **x,** as in Fig. 1C. In Fig. 4B, accurate RNA division rests on symmetrical outward-directed pairing (Fig. 3) by furrow-binding RNA **x**. More specific analysis of fatty acids awaits more data.

**Biology as Anthology.** Anthology means that evolution combines characters evolved separately, quickly assembling a fitter organism. Anthology rationalizes orderly evolution of favored qualities before a DNA genome, logically required to evolve the DNA genome itself (35). Mechanistically, evolution chooses a near-ideal subset from an entirely random reaction sequence (36). This is true, for example, of the genetic code, whose encodings for different amino acids were likely developed separately. Late code evolution slows dramatically for kinetic reasons (37), thereby hosting a long-lived crescendo (38) of varying codes closely resembling the Standard Genetic Code. One of these is selected as the Standard Genetic Code in the first organism that successfully spans diverse ecological conditions (39). Fig. 4 is particularly relevant, positing a plausible physical mechanism to combine ancient RNA functions, utilizing selected base pairings and shared bilayer affinity. Cell fusions would speed evolution (34), anthologizing protogenomic advances in an entire community of related microbes (33).

**Extant membrane RNAs?** This text combines a review of RNA-membrane interactions (**Introduction**) with a conjecture that they enable genetic tasks (**Discussion**). Existence of postulated reactions can be investigated with simple experiments, whose methods are established. The route from existing data (Fig. 2, Fig. 3) to plausible primordial RNA-membranes with biochemical and genetic functions (Fig. 4), or to modern RNA-membranes with biological roles (10), may be short indeed.

**RNA-membranes supply something indispensable.** Faster, more precise division is the necessity for evolutionary success for primitive microbes (40). The furrow of Fig. 4C promises just that, ensuring the success of the microorganism that controlled it. Thus, Earth's full biota can plausibly descend from microbes with Fig. 4's ancient division-promoting RNAs.

**Figure legends**

**Figure 1A-Cross-section of a bilayer lipid membrane**. Membranes discussed here, for example, can be composed of palmitic acid, a saturated 16-carbon fatty acid whose headgroups (blue) are a negatively charged carboxyl, and whose acyl groups (yellow) are a 15-carbon single-bonded chain of $-CH_2$, terminated by $-CH_3$. Modern phospholipid membranes contain phosphatidyl cholines, with headgroups (blue) of positively charged choline linked to negatively charged phosphate linked to glycerol; they are therefore zwitterionic and overall neutral with strong dipole moments. Hydrophobic acyl groups (yellow) esterified to glycerol can be palmitic acid plus oleic acid, an 18-carbon fatty acid kinked by one double bond at carbon 9.

**Figure 1B-A bilayer with bound RNA.** The RNA is simplified to a red linear backbone. The RNA is confined to the leaflet it initially contacted, as shown.

**Figure 1C-A distorted membrane-RNA.** A bilayer with bound RNAs, which in multiple copies can stabilize a bent membrane conformation. Leaflets A and B can contain lipids with particular shapes, which also favor bent bilayers. Though multiple active RNAs, as shown, are discussed, membranes can also be deformed by multiple uniform RNAs.

**Figure 2A-RNA10 sequence and activity.** A fold for the RNA10 113-mer (15) is shown, derived from a minimum free energy computation (41), with added structures from Rosetta (42) and reconciliation with measurements of inter-nucleotide flexibility (15). The structure is marked to show RNA9:RNA10 interaction sequences in pink lettering (15), the region replaced (blue marked arrows) by a tryptophan site in RNA10Trp (17), the region replaced by an arginine site in RNA10Arg (18), and the region of RNA10Arg doped and reselected for improved membrane and arginine binding. The sequence change after doped reselection (18) is also marked, in green lettering.

**Figure 2B-the RNA9 sequence.** A minimal free energy fold estimate (41) is shown.

**Figure 3A- Simplified graphics for RNA9 and RNA 10 structures.** RNA9 is in blue, RNA10 and functionalized RNA10 (17, 18) in red. Yellow numerals in larger loops are RNA names. Figure boundaries represent RNA backbones, and green arrowheads show their polarities. Black letters beside flattened loops are RNA sequences. Triangular intercalations are hypothetical functional domains having RNA functions **a**, **b**, **c**, **d** and **e**.

**Figure 3B- An RNA9:RNA10 structure that yields chains of arbitrary length (5).** The chain shown can be extended by an arbitrary number of RNA10:RNA9 dimers, symbolized by dots at the right.

**Figure 3C**- **An oligomeric RNA9:RNA10 structure that yields a stoichiometry of 2 RNA9: 1 RNA10 (5).**

**Figure 4A An RNA-membrane protogenome hypothesis.** The background gray sheet is the inner side of an RNA cell boundary. Blue regions are a rafted area in the boundary bilayer. Red objects are simplified RNA molecules using the notation of Fig. 3; functionalized RNA10-like RNAs with terminal loops having complementary sequences, and internal functional domains **a**, **b** and **f** or **m**, **n** and **o**; as in RNA10Trp (17) and RNA10Arg (18) structures (Fig. 2). Functional domains **x** and **y** resemble Figure 1C, and stabilize indentation (Fig. 4C) of the cell outer membrane (5).

**Figure 4B**- Protogenomic activity, but consistent with RNAs bound on a simpler, unrafted fatty acid membrane (9).

**Figure 4C**- A growing depression in the cell bounding membrane, ultimately capable of dividing the cell and its membrane. RNAs may either initiate a division furrow or reinforce a nascent furrow by joining it.

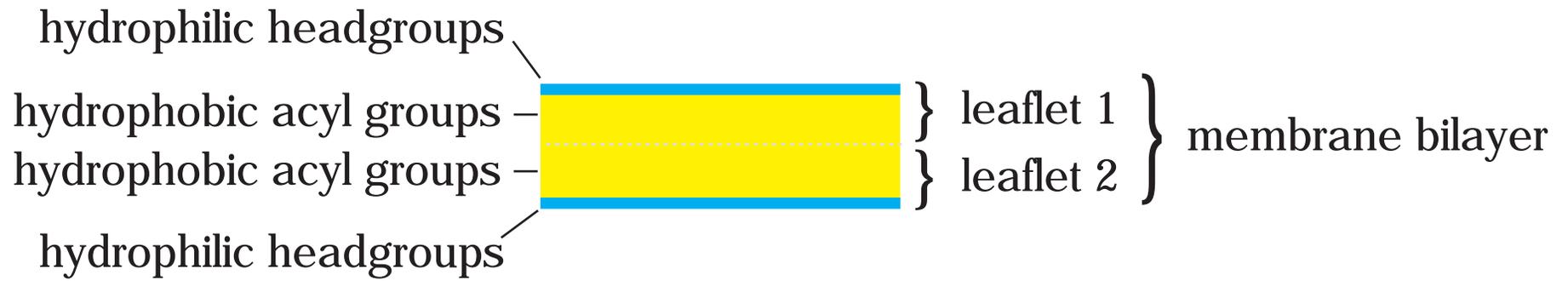

Figure 1A

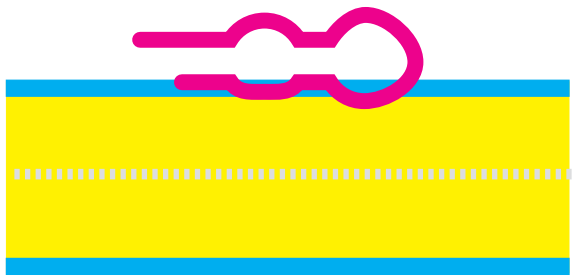

Figure 1B

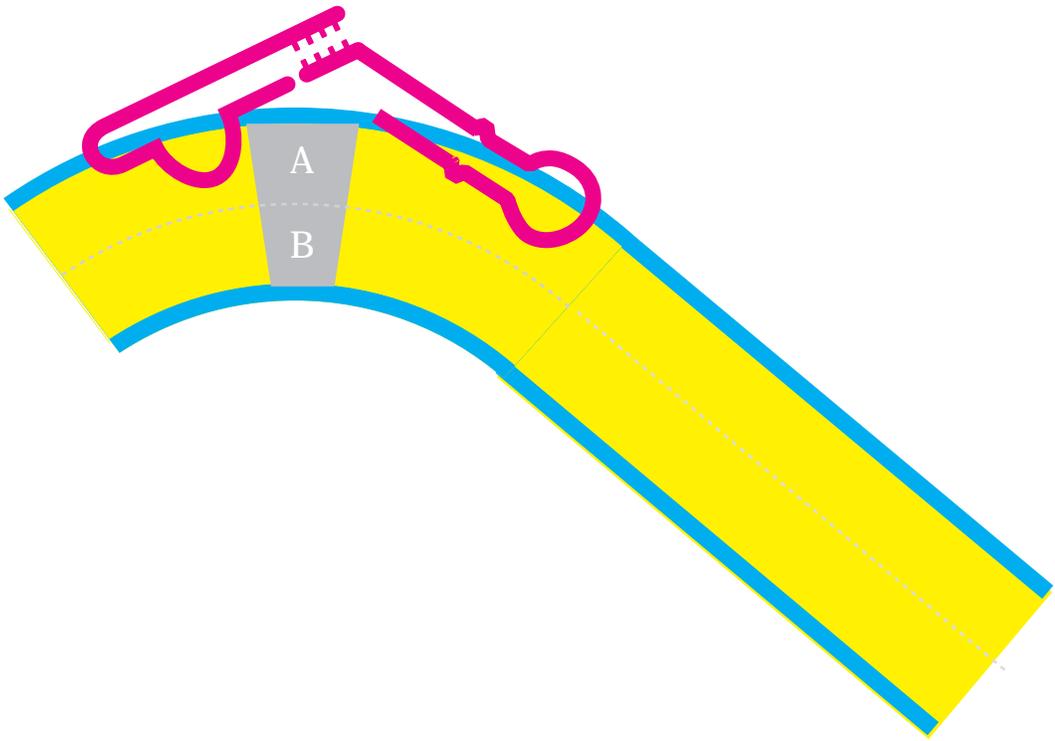

Figure 1C

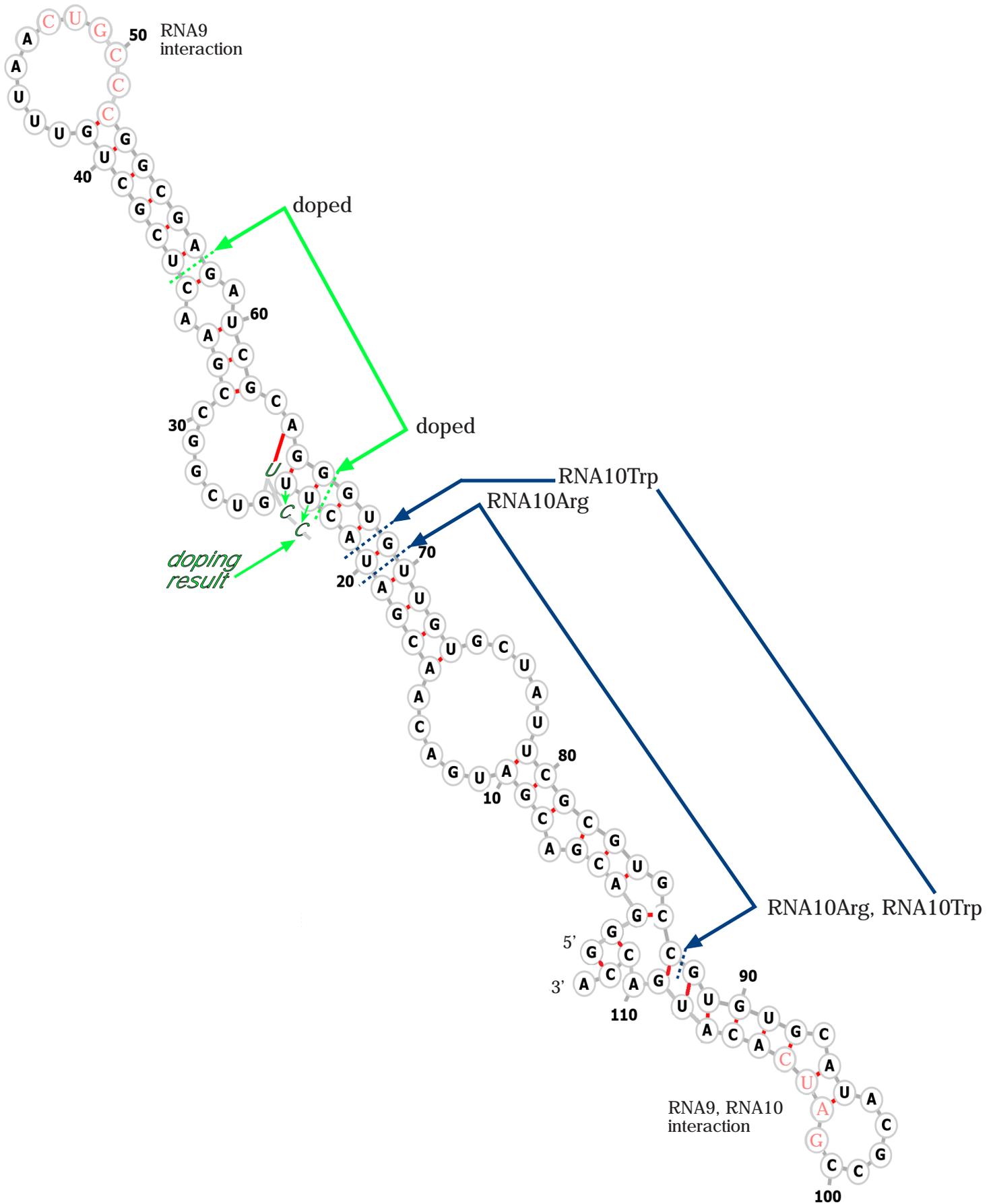
Figure 2A

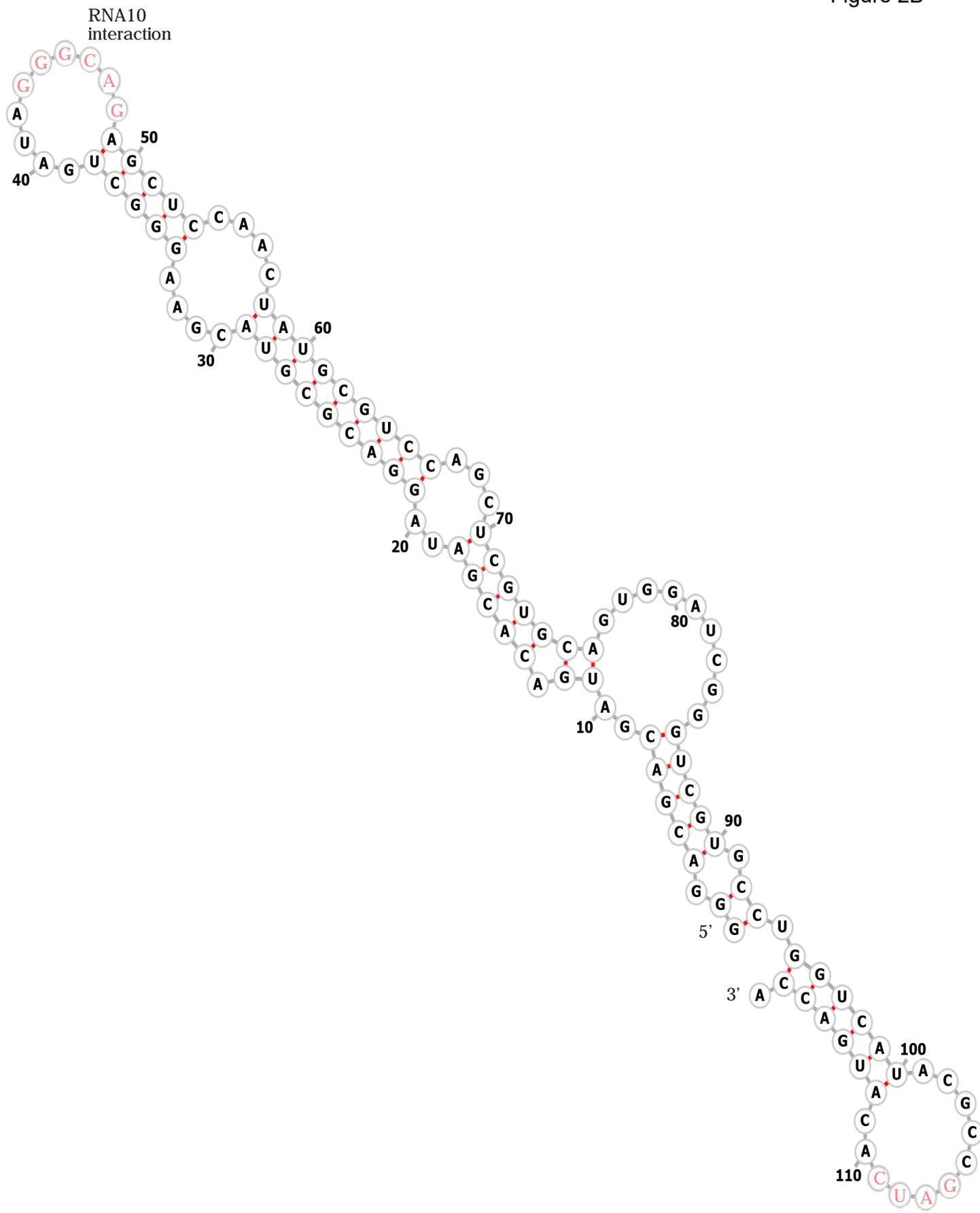

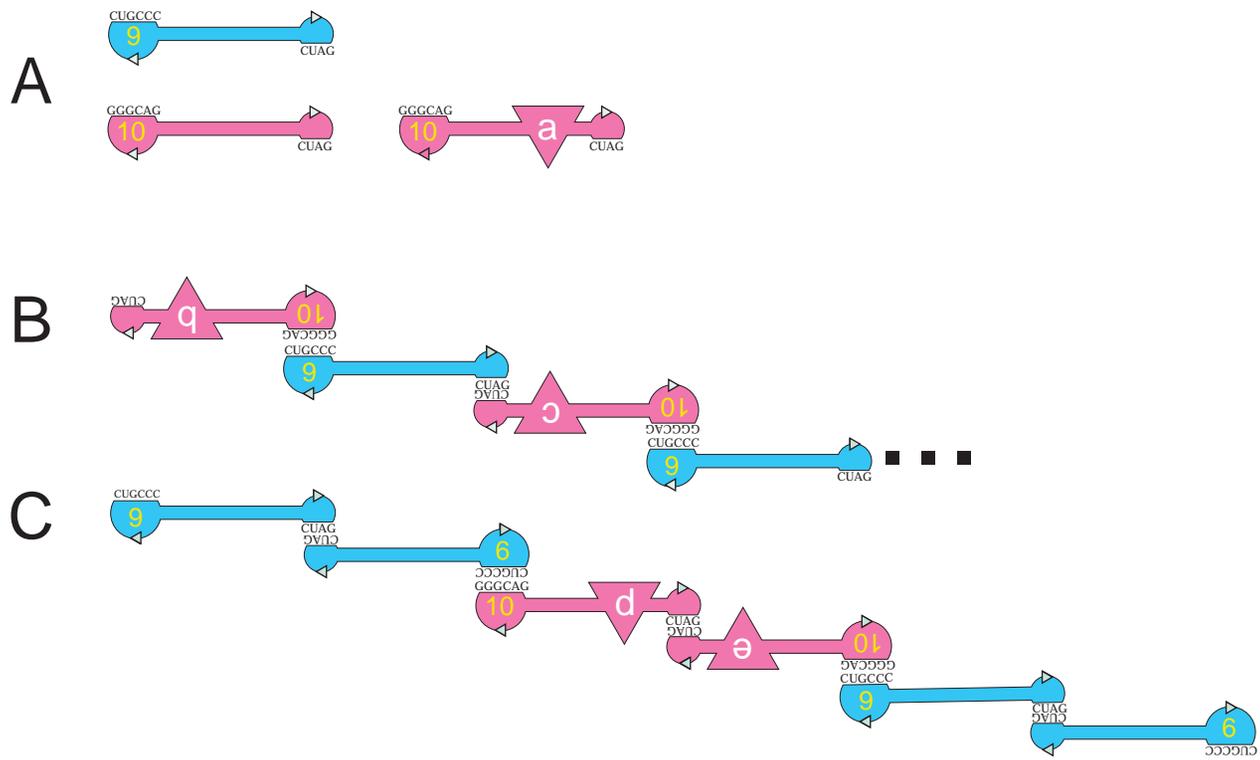

Figure 3

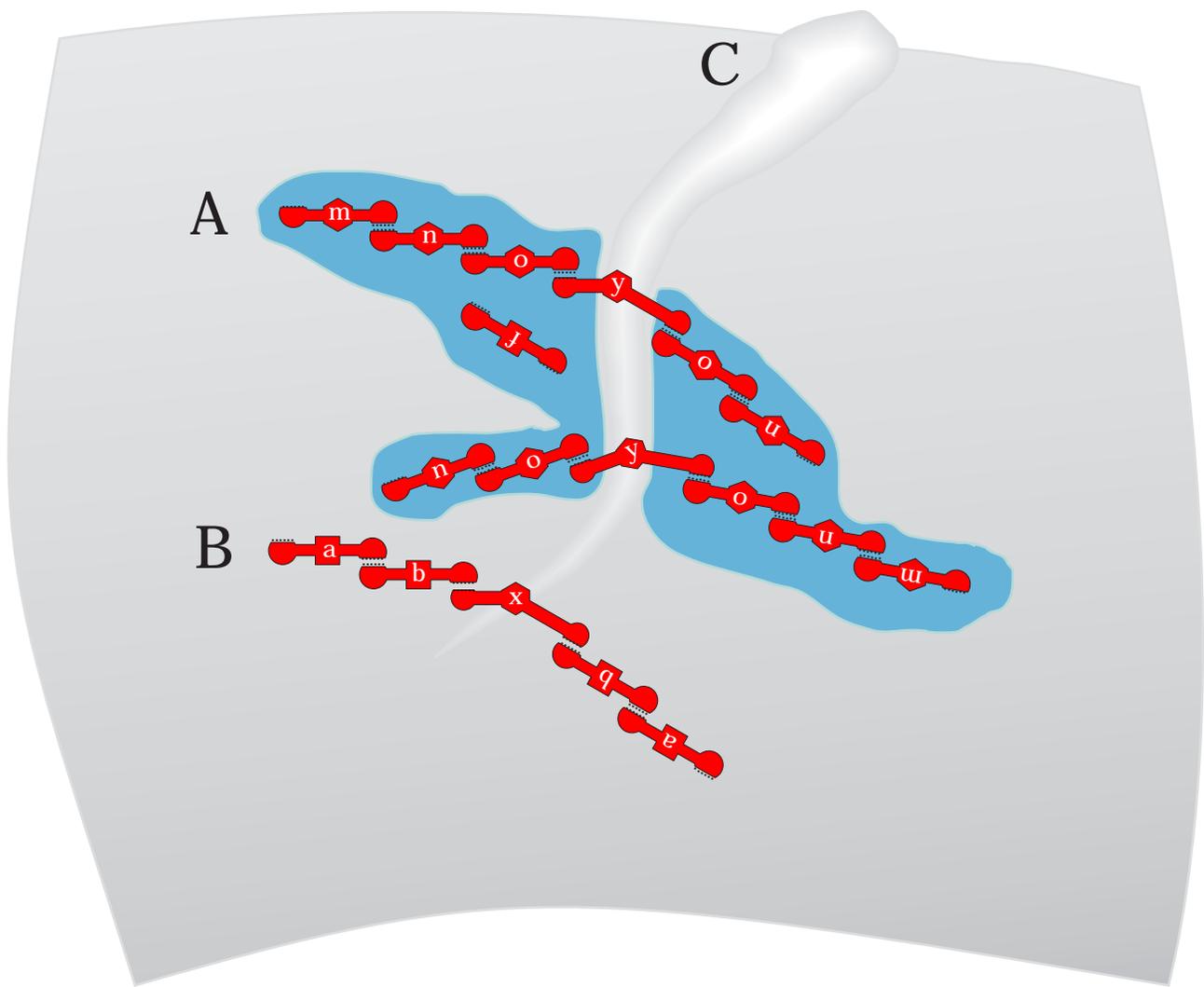

Figure 4